\documentclass[12pt,twoside,letterpaper]{article}
\usepackage[margin=20mm]{geometry}
\usepackage{upgreek} 
\newcommand{\micro}{\ensuremath{\upmu}}  
\newcommand{\PI}{\uppi}         
\begin{document}
\title{The influence of Earth rotation\\
in neutrino speed measurements\\
between CERN and the OPERA detector}
\author{Markus G. Kuhn\\
Computer Laboratory, University of Cambridge}
\date{20 October 2011}
\maketitle
\begin{abstract}
The OPERA neutrino experiment at the underground Gran Sasso Laboratory
recently reported, in arXiv:1109.4897v1, high-accuracy velocity
measurements of neutrinos from the CERN CNGS beam over the 730~km
distance between the two laboratories. This raised significant
interest, as the observed neutrinos appeared to arrive at the OPERA
detector about 60~ns (or equivalently 18 m) earlier than would have
been expected if they had traveled at the speed of light, with high
statistical significance. As the authors did not indicate whether and
how they took into account the Coriolis or Sagnac effect that Earth's rotation
has on the (southeastwards traveling) neutrinos, this brief note
quantifies this effect. It would explain a $2.2$~ns later
arrival time.
\end{abstract}

The speed of light in vacuum is $c=299\,792\,458$~m/s in inertial
frames of reference, that is, coordinate systems that do not rotate or
accelerate. However, in the ground-based neutrino-speed measurements
reported in \cite{measurement}, the emitter and detector rotate
eastwards with the Earth's crust, and the resulting Coriolis effect
(in optics also known as Sagnac effect) should be taken into account.

\newcommand{\cern}{a}
\newcommand{\opera}{b}
\newcommand{\cernv}{\mathbf{\cern}}
\newcommand{\operav}{\mathbf{\opera}}

The emitter coordinates $\mathbf{a}$ at CERN and the detector
coordinates $\mathbf{b}$ at OPERA are provided in Table 5 of OPERA
Public Note 132~\cite{opera132} as
\[
\cernv = 
\left(
\begin{array}{l}
\cern_\mathrm{X}\\
\cern_\mathrm{Y}\\
\cern_\mathrm{Z}
\end{array}
\right) = 
\left(
\begin{array}{r}
4\,394\,369.327~\mathrm{m} \\ 467\,747.795~\mathrm{m} \\ 4\,584\,236.112~\mathrm{m}
\end{array}
\right), \qquad
\operav =
\left(
\begin{array}{l}
\opera_\mathrm{X}\\
\opera_\mathrm{Y}\\
\opera_\mathrm{Z}
\end{array}
\right) = 
\left(
\begin{array}{r}
4\,582\,167.465~\mathrm{m} \\ 1\,106\,521.805~\mathrm{m} \\ 4\,283\,602.714~\mathrm{m}
\end{array}
\right).
\]
These are Cartesian coordinates in an Earth-centred Earth-fixed (ECEF)
coordinate system known as ETRF2000, which is commonly used for
geodetic work in Europe.\footnote{The origin of such geodetic XYZ
  coordinate systems is approximately the center of gravity of Earth,
  the length unit is 1~m, the Z axis points towards the North Pole,
  the X axis points towards the prime meridian and surfaces at
  longitude $0^\circ$ on the equator off the coast of West Africa, and
  the Y axis completes a right-handed coordinate system, surfacing at
  longitude $90^\circ$ east on the equator in the Indian Ocean.}

The distance
\[
||\operav-\cernv|| = \sqrt{(\opera_\mathrm{X}-\cern_\mathrm{X})^2+
                         (\opera_\mathrm{Y}-\cern_\mathrm{Y})^2+
                         (\opera_\mathrm{Z}-\cern_\mathrm{Z})^2}
 = 730\,534.610~\mathrm{m}
\]
appears to have formed in \cite[p.~10]{measurement} the basis for the
presented neutrino speed calculation, corrected only by local
equipment-related delays, distances and calibration parameters at each
end.

At the speed of light $c$, a particle travels that distance in about
\[
\Delta t = ||\operav-\cernv|| / c = 2.43680116~\mathrm{ms}.
\]

The Earth rotates eastwards around the Z axis, completing one
revolution per stellar day of about
$23~\mathrm{h}~56~\mathrm{min}~4~\mathrm{s} = 86\,164$~s, resulting in
an angular frequency of about
\[
\omega = 2\PI / 86\,164~\mathrm{s} =  72.921~\micro\mathrm{rad/s}.
\]
During the neutrino's time-of-flight $\Delta t$, Earth will
rotate by an angle of
\[
\phi = \omega\cdot \Delta t = 177.695~\mathrm{nrad}.
\]
We now define a new coordinate system IRF that is identical to
ETRF2000 at the time instant when the neutrino leaves the emitter
location, but does not rotate. For the distance calculation in IRF,
the emitter coordinates $\cernv$ remain the same as in ETRF2000.
However, the receiver coordinates $\operav$ at the time of arrival
change in IRF to
\[
\operav'= 
\left(
\begin{array}{ccc}
\cos\phi & -\sin\phi & 0 \\
\sin\phi & \cos\phi & 0 \\
0 & 0 & 1
\end{array}
\right)
\cdot \operav
\]
because ETRF2000 rotated eastwards around its Z axis during the
neutrino's flight duration $\Delta t$ by angle $\phi$, and with it the
OPERA detector, by $||\operav' - \operav|| = 0.838$~m.

In the non-rotating coordinate system IRF, the distance traveled by
the neutrino changes to
\[
||\operav' - \cernv|| = 730\,533.949~\mathrm{m} = ||\operav - \cernv|| +
0.661~\mathrm{m}.
\]
Therefore, without taking into account the Coriolis/Sagnac effect that
emerges when the neutrino is observed in a rotating coordinate system,
the neutrino would appear to arrive 0.661~m behind of where it should
be if it traveled at the speed of light, or 2.2~ns
late.\footnote{Thanks to John Field (CERN) for pointing out a sign
  error in the first version of this note (3 October 2011).} This
actually adds to the 60~ns measurement reported in \cite{measurement}.

\section*{Conclusion}

In the absence of the keywords ``inertial'', ``rotation'',
``Coriolis'', ``Sagnac'' in \cite{measurement,opera132}, it seemed
prudent to verify what impact a failure to perform the distance
calculation in a non-rotating coordinate system would have on the
observed time of flight. The result, a 2.2~ns later arrival, should be
taken into account\footnote{The authors of \cite{opera132} have since
  released Version 2 (10 October 2011), where section 7.3 now
  addresses this.}, but only strengthens the statistical significance
of the reported measurements.

\end{document}